\documentclass[12pt]{article}%
\usepackage{amsmath}
\usepackage{amsfonts}
\usepackage{amssymb}
\usepackage{graphicx}%
\providecommand{\U}[1]{\protect\rule{.1in}{.1in}}
\newcommand{\bfr}{\begin{flushright}}
\newcommand{\efr}{\end{flushright}}
\newcommand{\bc}{\begin{center}}
\newcommand{\ec}{\end{center}}
\newcommand{\ben}{\begin{enumerate}}
\newcommand{\een}{\end{enumerate}}

\newcommand{\be}{\begin{equation}}
\newcommand{\ee}{\end{equation}}
\newcommand{\ba}{\begin{array}}
\newcommand{\ea}{\end{array}}
\def\6{\partial}

\begin{document}

\title{\textbf{Entropic Law of Force, Emergent Gravity}
\\
\textbf{and the Uncertainty Principle} }
\author{ M. A. Santos$^{a}$\thanks{email: masantos@cce.ufes.br} 
and I. V. Vancea$^{b}$\thanks{email:ionvancea@ufrrj.br} 
\\$^{a}$\emph{{\small Departamento de F\'{\i}sica e Qu\'{\i}mica,}}
\emph{{\small Universidade Federal do Esp\'{\i}rito Santo (UFES),}}
\\
\emph{{\small Avenida Fernando Ferarri S/N - Goiabeiras, 29060-900 Vit\'{o}ria - ES, Brasil}}
\\
$^{b}$\emph{{\small Grupo de F{\'{\i}}sica Te\'{o}rica e Matem\'{a}tica
F\'{\i}sica, Departamento de F\'{\i}sica,}} 
\\
\emph{{\small Universidade Federal Rural do Rio de Janeiro (UFRRJ),}}
\\
\emph{{\small Cx. Postal 23851, BR 465 Km 7, 23890-000 Serop\'{e}dica - RJ,Brasil }}
}
\date{26 October 2011}
\maketitle

\thispagestyle{empty}


\abstract{The entropic formulation of the inertia and the gravity relies on quantum,
geometrical and informational arguments. The fact that the results are
completly classical is missleading. In this paper we argue that the entropic
formulation provides new insights into the quantum nature of the inertia and
the gravity. We use the entropic postulate to determine the quantum
uncertainty in the law of inertia and in the law of gravity in the Newtonian
Mechanics, the Special Relativity and in the General Relativity. 
These results are obtained by considering
the most general quantum property of the matter represented by the Uncertainty
Principle and by postulating an expression for the uncertainty of the entropy
such that: i) it is the simplest quantum generalization of the postulate of
the variation of the entropy and ii) it reduces to the variation of the
entropy in the absence of the uncertainty.}

\vfill


\newpage\pagestyle{plain} \pagenumbering{arabic}

\section{Introduction}

In a recent paper, Verlinde has put forward a very interesting proposal for
the origin of the inertia, the law of force, the law of gravity and the
General Relativity as emergent phenomena \cite{Verlinde:2010hp}. Acording to
this theory, the spacetime can be described as an information device made of
holographic surfaces (screens) on which the information about the physical
systems can be stored. The relevant information about the physical dynamics
can be recovered by analysing the variation of the information on the screens
and it is independent of the details of the particular theory used to describe
the physical system. The screens behave as streched horizons in the black hole
physics and define emergent holographic directions in which the spacetime
grow. The information on the screens is described by the information entropy
and, as the black hole entropy, it is encoded in a number of bits proportional
to the area of the screen \cite{Bekenstein:1973ur}. The total energy of the
degrees of freedom satisfy the equipartition theorem. Also, it is postulated
that the information entropy is maximized by the entropic forces that act
along the holographic directions and are defined by gradients of the entropy.
If one associates non-inertial frames to these forces, the corresponding
accelerated observers measure a redshifted information at the Unruh
temperature \cite{Unruh:1976db}%
\begin{equation}
T_{U}=\frac{\hbar a}{2\pi k_{B}c},\label{Unruh-temperature}%
\end{equation}
where $k_{B}$ is the Boltzmann's constant, $c$ is the speed of light and $a$
is the acceleration. For a particle of mass $m$, it is postulated in
\cite{Verlinde:2010hp} that the variation of the entropy along the holographic
direction be linear in the separation $\Delta x$ between the particle and the
screen%
\begin{equation}
\Delta S=2\pi k_{B}\frac{\Delta x}{l_{c}},\label{Verlinde-ansatze}%
\end{equation}
where $l_{c}=\hbar/mc$ is the Compton length. Thus, the Compton length defines
the units of the entropy change. The variation of the entropy $\Delta S$
corresponds to an arbitrary variation of the energy $\Delta W$ at the
thermodynamical equilibrium (with $T=T_{U}$) which, at its turn, can be
interpreted as the result of a macroscopic force $F$ acting on the particle
over the distance $\Delta x$%
\begin{equation}
\Delta W=F\Delta x=T\Delta S.\label{energy-force-entropy}%
\end{equation}
The above assumptions lead to the law of force for $m$ and when applied to a
spherical screen around a mass $M$ they deliver the Newton's law of gravity.
Also, when properly generalized, the postulates allow one to obtain the
Einstein's equations \cite{{Verlinde:2010hp}}\footnote{The Einstein's
equations have been obtained before from the thermodynamical considerations in
\cite{Jacobson:1995ab}. A very interesting proposal for the gravitational
entropy and its connection with the equipartition of the microscopic horizon
degrees of freedom in the context of the emergent gravity was made in
\cite{Padmanabhan:2003pk,Padmanabhan:2009kr}.}.

Several features and consequences of the entropic postulate have already been
explored. The compatibility between the entropic gravity and the loop quantum
gravity was proved in \cite{Smolin:2010kk}. Applications to cosmology,
Friedmann's equations and D-brane cosmology were developed in
\cite{Shu:2010nv,Cai:2010hk,Wang:2010jm,Wei:2010ww,Ling:2010zc}. The
connection between the entropic gravity and the black holes was explored in
\cite{Caravelli:2010be,Myung:2010jv,Liu:2010na,Kwon:2010km,Tian:2010uy}. A
modification of the entropic force to include the deviation from the
equipartition theory was proposed in \cite{Gao:2010fw}. The derivation of a
dark energy term from the entropic gravity was given in \cite{Li:2010cj}.
Further relations with the thermodynamics were discussed in
\cite{Zhao:2010qw,Zhang:2010hi,Culetu:2010iq}. A speculation about the
possible interpretation of the Coulomb force as an entropic force was done in
\cite{Wang:2010px}. The realisation of the information entropy in terms of
light was investigated in \cite{Pesci:2010un}. Other gometrical, field
theoretical and informational aspects of the theory are presented in
\cite{Cai:2010sz,KowalskiGlikman:2010ms,Lee:2010bg,Piazza:2009zh,Konoplya:2010ak}.

In the above construction, the holographic principle
\cite{'tHooft:1993gx,Susskind:1994vu} and the information entropy
\cite{Bekenstein:1973ur} play fundamental roles in establishing the entropic
nature of the inertia and of the law of force. On the other hand, since the
Planck constant cancells out in the law of force and the gravitational force,
it seems that the quantum effects are irrelevant to the inertia and gravity
despite the quantum nature of the key concepts from the relations
(\ref{Unruh-temperature}) and (\ref{Verlinde-ansatze}). Thus, one may ask the
important question whether there are quantum corrections to the laws of
Mechanics derivable from the entropic postulate. The aim of this paper is to
show that this questions has a positive answer if one takes into account the
quantum structure of matter in the most general form of the Uncertainty Principle.

The paper is organized as follows. In Section 2 we briefly review the Verlinde's postulate.
Then we use it to derive the uncertainty in the entropy of the holographic screen if
the particle is subjected to the Uncertainty Principle. We also argue that the uncertainty
in the entropy should imply an uncertainty in the adiabatic force which can be interpreted
as a quantum correction to it. In Section 3 we discuss the same situation in the
Special Relativity as well as in the General Relativity. The last section is devoted to some
discussions.

\section{Non-Relativistic Entropic Force and the Uncertainty Principle}

Consider a system composed by a holographic screen $\mathcal{S}$ and a quantum
test particle of mass $m$. Due to the quantum nature of the particle, the
separation between $S$ and $m$ is determined only up to the uncertainty
$\delta x$ that satisfies the Heiseberg's relation
\begin{equation}
\delta x\delta p\geq\frac{\hbar}{2},\label{Heisenberg}%
\end{equation}
where the uncertainty in the momentum along the transverse direction to
$\mathcal{S}$ is $\delta p$. If the particle were classical and $\delta x$
were zero, the variation of the entropy on $\mathcal{S}$ would be given by
(\ref{Verlinde-ansatze}) for any separation $\Delta x$. Since the test
particle is quantum and there is an uncertainty of its position and momentum,
an uncertainty of the entropy on the screen is expected. The uncertainty of
$S$ can be related to the uncertainty in the total energy by the first law of
the thermodynamics
\begin{equation}
\delta W=T\delta S.\label{uncertainty-energy-force-entropy}%
\end{equation}
The above equality holds at thermodynamical equilibrium at the temperature
$T$. The reason for the uncertainty of the energy is that both the kinetic
energy $K$ and the potential energy $V$ (which is the relative energy of the
particle with respect to the screen) are uncertain due to the uncertainty of
the position and momentum of the test particle. Thus, if the quantum particle
is non-relativistic, the relation (\ref{uncertainty-energy-force-entropy})
takes the following form
\begin{equation}
\frac{p}{m}\delta p+F\delta x=T\delta S.\label{energy-force-entropy-1}%
\end{equation}
We note that it is possible to add an extra term to the l. h. s. of the
relation (\ref{energy-force-entropy}) of the form $\delta Fx$. Such a term can
be produced by the variation of $\delta p$ in time. If that is fixed then
$\delta F=0$.

The relation (\ref{energy-force-entropy-1}) can be used to evaluate the
uncertainty of the entropy. Indeed, by inspecting
(\ref{energy-force-entropy-1}), one can see that the main difference between
the entropy function of a classical particle given in \cite{Verlinde:2010hp}
and the quantum particle is the dependence of the entropy on $Fx$ \emph{and}
$K$ for the latter, i. e. $S(U+K+Fx)$ for the quantum system. Moreover, the
relation (\ref{energy-force-entropy-1}) suggests that $\delta S$ depends not
only on the undeterminacy in the momentum but also on its value. By taking
into account all these considerations and the relation (\ref{Verlinde-ansatze}%
) for the classical particle, we are proposing the following relation for the
uncertainty of the entropy
\begin{equation}
\delta S=2\pi k_{B}\left(  \frac{\delta x}{l_{c}}+\frac{p\delta p}{m^{2}c^{2}%
}\right)  .\label{ansatze-entropy-1}%
\end{equation}
The denominators in the above relation result from dimensional considerations.
In the classical case, the uncertainty of the momentum of particle is zero and
the Heisenberg's relation does not apply. Then by interpreting $\delta x$ as
the separation $\Delta x$, the relation (\ref{ansatze-entropy-1}) reduces to
the Verlinde's ansatz (\ref{Verlinde-ansatze}). By using
(\ref{Unruh-temperature}), (\ref{ansatze-entropy-1}) and the estimate $\delta
p=\hbar(2\delta x)^{-1}$ from (\ref{Heisenberg}) in the equation
(\ref{energy-force-entropy-1}) one obtains the following relation
\begin{equation}
F(\delta)= ma + \frac{\hbar p}{2m}\left(  \frac{\hbar a}{mc^{3}}-1\right)  \delta x^{-2}.
\label{force-uncertainty-1}
\end{equation}
The above equation represents the entropic force acting on the quantum
particle subjected to the uncertainty principle. The classical limit is
obtained by taking simultaneously $\delta p=0$ and $\hbar\rightarrow0$ in the
above estimate, which renders the r. h. s. of the relation
(\ref{energy-force-entropy-1}) zero. Thus, the above equation represents a
generalization of the law of force to the Quantum Mechanics. If one defines
the uncertainty of the force as%
\begin{equation}
\delta F=F(\delta)-F,\label{uncertainty-force}%
\end{equation}
the following uncertainty relation holds%
\begin{equation}
\delta F\delta x^{2}\geq\frac{\hbar p}{m}\left(  \frac{\hbar a}{mc^{3}%
}-1\right)  .\label{uncertainty-force-position-1}%
\end{equation}
Note that in the relation (\ref{ansatze-entropy-1}) the uncertainty of $S$
depends not only on $\delta x$ and $\delta p$ but also on the momentum $p$.
There is no other argument for this assumption than the form of the l. h. s.
of the relation (\ref{energy-force-entropy-1}). In principle, one could relax
this condition and assume instead that $\delta S$ depends only on $\delta x$
and $\delta p$. In this case, the following relation is suitable for the
definition of the uncertainty of the entropy%
\begin{equation}
\delta S=2\pi k_{B}\left(  \frac{\delta x}{l_{c}}+\frac{\delta p}{mc}\right)
.\label{ansatze-entropy-2}%
\end{equation}
By performing the same steps as above, one obtains the following
generalization of the law of force%
\begin{equation}
F(\delta)=ma+\frac{\hbar}{2m}\left(  \frac{\hbar a}{c^{2}}-p\right)  \delta
x^{-2},\label{force-uncertainty-2}%
\end{equation}
with the corresponding uncertainty relation%
\begin{equation}
\delta F\delta x^{2}\geq\frac{\hbar}{2m}\left(  \frac{\hbar a}{c^{2}%
}-p\right)  .\label{uncertainty-force-position-2}%
\end{equation}

The above analysis shows that the postulate of the entropic force can be used
to obtain information about quantum corrections to the law of inertia and the
law of force, thus establishing a connection between the classical and the
quantum concepts. By particularizing these ideas, one can obtain quantum
corrections to the law of gravity. To this end, we consider a spherical screen
that encloses a mass $M$. Following \cite{Verlinde:2010hp}, we assume that the
information on the screen is encoded in a number $N$ of bits proportional to
the area of the sphere%
\begin{equation}
N =\frac{Ac^{3}}{G\hbar},\label{bits-area}%
\end{equation}
where $G$ is the Newton's constant. This information can be realized in terms
of a classical system with $N$ degrees of freedom and the energy cost $E$.
Since in the absence of any other matter the information is exclusively about
the mass $M$, the energy cost equals $Mc^{2}$ in the reference frame in which
$M$ is at rest. Also, the $N$ classical degrees of freedom being identical,
one can assume that the equipartition law holds at the thermodynamical
equilibrium%
\begin{equation}
E=\frac{1}{2}Nk_{B}T.\label{equipartition-theorem}%
\end{equation}
Next, one considers a quantum particle $m$ outside the sphere. From the
Heisenberg's principle, the uncertainty in its position and momentum leads to
an uncertainty in the information on the sphere as argued in the general case.
Then, by applying the results from (\ref{force-uncertainty-1}), one can see
that%
\begin{equation}
F_{g}(\delta)=G\frac{Mm}{R^{2}}+\frac{\hbar p}{2m}\left(  G\frac{\hbar
M}{R^{2}mc^{3}}-1\right)  \delta x^{-2}.\label{uncertainty-gravity-1}%
\end{equation}
The above equation represents the quantum gravitational force derived from the
entropic force postulate and the Heisenberg's principle. In the classical
limit, it reproduces the Newton's law of gravity. If the uncertainty in the
entropy does not depend on the momentum of $m$, the relation
(\ref{force-uncertainty-2}) should be used instead of
(\ref{force-uncertainty-1}) and the corresponding quantum gravitational force
takes the following form
\begin{equation}
F_{g}(\delta)=G\frac{Mm}{R^{2}}+\frac{\hbar}{2m}\left(  G\frac{\hbar M}%
{R^{2}c^{2}}-p\right)  \delta x^{-2}.\label{uncertainty-gravity-2}%
\end{equation}
Similar force-position uncertainty relations to the ones given in
(\ref{uncertainty-force-position-1}) and (\ref{uncertainty-force-position-2})
can be derived for the gravitational force, too. It is important to note that
even if the separation between $\mathcal{S}$ and $m$ is along the normal
direction to $S$, the delocalization of the particle is along $x$, $y$ and
$z$. Furthermore, there is an uncertainty of the energy within the bounds of
the inverse of the observation time. The generalization of the above relations
to these cases is straightforward in the non-relativistic theory.

\section{Relativistic Entropic Force and the Uncertainty Principle}

We can try to inferr the uncertainty of the entropy in the relativistic case
from the same heuristics. The basic relations of the relativistic dynamics and
the equation (\ref{uncertainty-energy-force-entropy}) lead to the following
relationship among the uncertainties in $x$, $p$ and $S_{r}$%
\begin{equation}
\frac{p\delta p}{m\gamma^{2}}+F\delta x=T\delta S_{r}%
,\label{energy-force-entropy-relat}%
\end{equation}
where $p=m\gamma v$ is the relativistic momentum of the test particle and
$\gamma$ is the Lorentz factor of the inertial frame of the particle
relatively to the inertial frame associated to the screen. By using the same
arguments as before, one is led to postulate the following uncertainty
relation in the entropy on the holographic screen
\begin{equation}
\delta {S'}_{r}=2\pi k_{B}\left(  \frac{\delta x}{l_{c}}+\frac{p\delta p}%
{m^{2}c^{2}\gamma^{2}}\right).
\label{ansatze-entropy-rel-1}%
\end{equation}
The above equation generalizes correctly the relation (\ref{ansatze-entropy-1}%
) as can be seen by taking the non-relativistic limit $v/c\rightarrow0$ while
keeping $c$ constant, where $v$ is the relative velocity of the two inertial
frames. In this limit, the denominator of the second term from the r. h. s. of
the equation (\ref{ansatze-entropy-1}) is recovered along with the classical
quantities.
Despite of that, the equation (\ref{ansatze-entropy-rel-1}) \emph{does not} lead to
the relativistic generalization of the $F(\delta)$ from the equation (\ref{force-uncertainty-1})
as one would naively expect, but instead one obtains the following equation%
\begin{equation}
{F'}_{r}(\delta)= ma + 
\frac{\hbar p}{2m\gamma^{2}}\left(  \frac{\hbar a}{mc^{3}}-1\right) \delta x^{-2}.
\label{not-rel-inertia-1}
\end{equation}
In order to obtain the uncertainty in the relativistic force acting on $m$ due to the uncertainty in the entropy on the screen, 
one should modify not only the ansatze given by the equation (\ref{ansatze-entropy-rel-1}) but also the 
Verlinde's postulate from (\ref{Verlinde-ansatze}). We propose the following relation for the relativistic uncertainty in the entropy   
\begin{equation}
\delta S_{r} =2 \pi k_{B} \left( \gamma^3 \frac{\delta x}{l_{c}} + \frac{p\delta p}{m^{2}c^{2}\gamma^{2}}\right).
\label{ansatze-entropy-rel-2}
\end{equation}
The first term in the above equation describes the variation of the entropy of the screen for the non quantum particle. We see that
it gains an $\gamma^3$ factor which is hard to be foresought without comparing the entropy against the force. From (\ref{ansatze-entropy-rel-2})
one can easily obtain
\begin{equation}
F_{r}(\delta)= m\gamma^3 a + 
\frac{\hbar p}{2m\gamma^{2}}\left(  \frac{\hbar a}{mc^{3}}-1\right) \delta x^{-2}.
\label{not-rel-inertia-2}
\end{equation}
The $F_{r}(\delta)$ represents the correct generalization of the equation(\ref{force-uncertainty-1}) to the relativistic case. Similarly, one can
obtain the corresponding generalizations of the equations (\ref{ansatze-entropy-2}) and (\ref{force-uncertainty-2}) which are left as an exercise.

Now let us extend these results to the General Relativity. The setting to
discuss the effect of the Uncertainty Principle in the entropic general
relativity is the one used to derive the entropic relativistic force from
\cite{Verlinde:2010hp}. We consider a static background with a global timelike
Killing vector $\xi^{a}$ which is necessary to define the temperature and the
entropy variation and which determines the relativistic generalization of the
Newton's potential $\phi$. In this background, the particle of rest mass $m$
is accelerated with the acceleration
\begin{equation}
a^{b}=u^{a}\nabla_{a}u^{b}=-\nabla^{b}\phi, \label{relativistic-acceleration}%
\end{equation}
where $u^{a}$ is the relativistic velocity of the particle (see
\cite{Wald:1984rg}). Whenever the test particle is subjected to an
acceleration, a relativistic force can be defined as $a^{b}=f^{b}/m$. In the
entropic general relativity, the acceleration is produced by the entropy
gradient along the normal direction to the holographic screen $\mathcal{S}$
located at surfaces of constant redshift defined by $e^{\phi/c^{2}}$. This
entropic acceleration should be related to an entropic force as mentioned
above. The unit vector pointing in the normal direction to $\mathcal{S}$ is
denoted by $N^{a}$. As in the non-relativistic case, the information entropy
on the screen can be related to the mechanical entropy of the system at local
thermodynamical equilibrium with the local temperature
\begin{equation}
T=\frac{\hbar}{2\pi k_{B}c}e^{\frac{\phi}{c^{2}}}N^{a}\nabla_{a}\phi,
\label{local-temperature}%
\end{equation}
which corresponds to the $T_{U}$ in the non-relativistic case. The local
variation of the entropy is defined along the normal direction to the screen
and it is postulated to have the following value for the Compton length
\cite{Verlinde:2010hp}
\begin{equation}
\nabla_{a}S=-\frac{2\pi k_{B}}{l_{c}}N_{a}, \label{local-variation-entropy}%
\end{equation}
where the minus sign is due to the crossing of the holographic surface from
outside to inside. As in the non-relativistic case defined by the equation
(\ref{Verlinde-ansatze}), the above relation defines the units of the
variation of the entropy.

Now let us assume that the test particle has a quantum mechanical description
in some local neighbourhood of $\mathcal{S}$ (for at least short times and
small spacelike volumes). Then it is subjected to the local relativistic
generalization of the Heisenberg's principle $\delta x_{a} \delta p_{a}
\geq\hbar/2$. In this case, an observer that moves with the velocity $v^{a}$
in the neighbourhood of $\mathcal{S}$ and $m$ will register an uncertainty of
the local information on the screen $\delta S$ related to the uncertainty of
the entropy by the equation
\begin{equation}
\delta W_{v} = -v^{a} \delta p_{a} + F_{a} \delta x^{a} = T \delta S.
\label{uncertainty-energy-force-entropy-relativistic}%
\end{equation}
Here, the index $v$ of the uncertainty of the energy means that the
uncertainty is observer dependent. The relation
(\ref{uncertainty-energy-force-entropy-relativistic}) represents the
relativistic generalization of (\ref{uncertainty-energy-force-entropy}) from
the non-relativistic case. If we interpret
(\ref{uncertainty-energy-force-entropy-relativistic}) as being the definition
of the uncertainty of S as we did in the non-relativistic case, we conclude
that the uncertainty in the entropy should be observer dependent. Then, the
simplest relation that can be postulate for the uncertainty of the entropy is
\begin{equation}
\delta S_{v} = - 2 \pi k_{B} \left(  \frac{N_{a} \delta x^{a}}{l_{c}} +
\frac{m v_{a} \delta p^{a}}{p^{2}} \right)  .
\label{ansatze-entropy-relativity}%
\end{equation}
By plugging the equations (\ref{local-temperature}) and
(\ref{ansatze-entropy-relativity}) into
(\ref{uncertainty-energy-force-entropy-relativistic}) and using the
uncertainty estimate $\delta p_{a} = \frac{\hbar}{2}\delta x^{-1}_{a}$ where
$\delta x^{-1}_{a} = \delta x_{a} / \delta x^{2} $, one can show that the
entropic relativistic force when the uncertainty is taken into account has the
following form
\begin{equation}
F^{a}_{v} (\delta) = - m e^{\frac{\phi}{c^{2}}} \nabla^{a} \left( \frac{\phi}{c^2} \right) -
\frac{\hbar}{2} \left[  1 + \frac{\hbar }{m c^3} e^{\frac{\phi}{c^{2}}} N^{b} \nabla_{b} \left( \frac{\phi}{c^2} \right) \right]  v^{a} \delta x^{-2}.
\label{uncertainty-gravity-relativity}
\end{equation}
The Newtonian limit is defined as usual by taking the weak field approximation and the slow
moving particle limit
\begin{eqnarray}
g_{ab} & = & \eta_{ab} + h_{ab},~~ h_{ab} << 1,~~ h_{00} = 2 \frac{\phi}{c^2},
\label{newton-limit}\\
\frac{d x^0}{d \tau} & \cong & 1,~~ \left| \frac{d x^{i}}{d \tau} \right| << \left| \frac{d x^0}{d \tau} \right|.
\label{slow-part-limit}
\end{eqnarray}
Then for the flat space defined by the condition $\phi / c^2 << 1 $ the equation (\ref{uncertainty-gravity-relativity}) reproduces the equation (\ref{force-uncertainty-1}). It is important to remark
that the General Relativity seems to put a stronger constraint on the
uncertainty of the entropy than the Newtonian mechanics. Indeed, the postulate
given in the relation (\ref{ansatze-entropy-relativity}) is the simplest
linear dependence of the uncertainty of the entropy on $\delta x^{a}$ and
$\delta p^{a}$. The point is that it also contains the velocity of the
observer $v^{a}$ as a consequence of the second equality from
(\ref{uncertainty-energy-force-entropy-relativistic}). Formally, one could
replace $v^{a}$ by a different vector. But then the equality could not hold in
the variable $v^{a}$. Thus, the simplest linear dependence on $\delta x^{a}$
and $\delta p^{a}$ forces the presence of $v^{a}$ in $\delta S$ as a logical
and physical requirement of the theory. An interesting consequence of this is
that the observer with $v^{a} = 0$ does not see any uncertainty in the
gravitational force. The similar constraint in the classical case (the
dependence of $\delta S$ on $p$), does not seem so strong and allows two
different interpretations of the uncertainty of the entropy as discussed before.

\section{Conclusions and Discussions}

To conclude, the postulate of the entropic force represents a powerfull tool
for investigating the quantum nature of the gravitational phenomena. This
feature stems from its quantum, geometric and informational foundations. The
fact that its most impressive results, the new proposal about the nature of
the inertia, the Newtonian gravity and the General Relativity obtained in
\cite{Verlinde:2010hp} are all classical hides the predictive power of the
entropic postulate at the quantum level. In this paper, we have shown that by
considering the simplest and the most fundamental quantum property of the
matter, the Uncertainty Principle, we are led to new insights in the quantum
structure of the inertia and the gravity in both the non-relativistic and the
relativistic theories, in the form of the quantum undeterminacy of forces.
Generalizations of the results presented here and of the Uncertainty Principle
can be found in \cite{Ghosh:2010hz,Banerjee:2010yd,Banerjee:2010ye,Banerjee:2010sd}.

Our results have been obtained by postulating the expressions
(\ref{ansatze-entropy-1}) and (\ref{ansatze-entropy-2}) for the undeterminacy
of the entropy in the non-relativistic theory. With these relations given, we have 
argued that the correct generalization of the uncertainty of the entropy on the screen
in the Special Relativity is given by the equation \ref{ansatze-entropy-rel-2}) while 
in the General Relativity is the equation (\ref{ansatze-entropy-relativity}) that should be considered for that purpose. 
The criteria used to establish these relations were the simplicity and the generalization
of the classical postulate from \cite{Verlinde:2010hp} that is recovered in
the classical limit of the quantum theory. The guide to establish the
dependence of $\delta S$ on $\delta x$ and $\delta p$ is the dependence of
$\delta W$ on $\delta x$ and $\delta p$, since $\delta W$ equates to $\delta
S$ through the first law of the thermodynamics. Thus, interpreting $\delta x$
and $\delta p$ as variables strongly constraints the form of $\delta S$. (For
example, terms of the form $\delta F = 0$ are not allowed.) However, the above
criteria alone are not enough to discern between the relations
(\ref{ansatze-entropy-1}) and (\ref{ansatze-entropy-2}). 

This situation is improved in the General Relativity, where one can consider the velocity of
the observer $v^{a}$ as an extra variable and one can require that it appear
explicitly in the $\delta S$ as we have done in
(\ref{ansatze-entropy-relativity}). Indeed, since the uncertainty of the
energy $\delta W_{v}$ depends on the observer, it does not seem natural to
require that $\delta S$ be observer independent as the first law makes these
two quantities proportional to each other. By taking this point of view
together with the previous criteria, the undeterminacy of the entropy is
uniquely fixed to have the form given by the relation
(\ref{ansatze-entropy-relativity}). If the above argument is ignored, then
$\delta p^{a}$ can enter $\delta S$ through different terms from which the
most simple ones have the form $p_{a} \delta p^{a} / p^{2} $ and $N_{a} \delta
p^{a} / m c$. In these cases, $\delta F^{a}_{v}$ takes the form
\begin{align}
\delta F^{a}_{v}  &  = \frac{\hbar}{2} \left[  v^{a} - \frac{\hbar p^{a}}{m^2 c }
e^{\frac{\phi}{c^{2}}} N^{b} \nabla_{b} \left( \frac{\phi}{c^2} \right)  \right]
\delta x^{-2} ,
\label{F-p-relativistic}\\
\delta F^{a}_{v}  &  = \frac{\hbar}{2} \left[  v^{a} - \frac{\hbar}{m }
e^{\frac{\phi}{c^{2}}} N^{a} N^{b} \nabla_{b} \left( \frac{\phi}{c^2} \right) \right]  \delta x^{-2}.
\label{F-N-relativistic}
\end{align}
Note that (\ref{F-p-relativistic}) and (\ref{F-N-relativistic}) are zero in
the classical limit. However, in both cases the undeterminacy registered by
the observer with $v^{a} = 0$ is no longer zero. 

An interesting generalization of the results presented in this paper concerns
the equipartition theorem. The derivation of the uncertainty of the law of
gravity depends on the validity of the equipartition theorem as in
\cite{Verlinde:2010hp}. If qubits are considered instead of bits, the
equipartition theorem can still be applied since a qubit can be realized in
terms of a physical system with one degree of freedom. However, it is well
known that the equipartition law fails for many quantum systems due to frozen
modes that can exist at low temperatures. Since our arguments rely on the
quantum nature of the test particle, it would be interesting to discuss the
alternatives to the equipartition theorem which, in general, are model dependent.

\noindent\textbf{Acknowledgements} .

We thank L. Holender and M. T. D. Orlando for stimulating discussions. Also, we acknowledge an anonymous referee
for suggesting important improvements. I. V. V. would like to acknowledge M. T. D. Orlando for hospitality at 
DFQ-CCE-UFES where this work was completed.



\end{document}